\documentclass[aps,prb,final,twocolumn,superscriptaddress]{revtex4-1}
\usepackage{amsmath}
\usepackage{graphicx}
\usepackage{epsfig}

\begin{document}

%\preprint{}

\title{Stability of the skyrmion lattice near the critical temperature in 
cubic helimagnets}


\author{Victor Laliena}
\email[]{laliena@unizar.es}
%\thanks{}
\affiliation{Instituto de Ciencia de Materiales de Arag\'on 
(CSIC -- Universidad de Zaragoza) and \\ 
Departamento de F\'{\i}sica de la Materia Condensada, Universidad de Zaragoza \\
C/Pedro Cerbuna 12, 50009 Zaragoza, Spain}
\author{Germ\'an Albalate}
%\thanks{}
\affiliation{Instituto de Ciencia de Materiales de Arag\'on 
(CSIC -- Universidad de Zaragoza) and \\ 
Departamento de F\'{\i}sica de la Materia Condensada, Universidad de Zaragoza \\
C/Pedro Cerbuna 12, 50009 Zaragoza, Spain}
\author{Javier Campo}
\email[]{javier.campo@csic.es}
%\thanks{}
\affiliation{Instituto de Ciencia de Materiales de Arag\'on 
(CSIC -- Universidad de Zaragoza) and \\ 
Departamento de F\'{\i}sica de la Materia Condensada, Universidad de Zaragoza \\
C/Pedro Cerbuna 12, 50009 Zaragoza, Spain}


%\date{\today}
\date{July 19, 2018}

\begin{abstract}
The phase diagram of cubic helimagnets near the critical temperature is obtained from
a Landau-Ginzburg model, including fluctuations to gaussian level. The free energy is
evaluated via a saddle point expansion around the local minima of the Landau-Ginzburg
functional. The local minima are computed by solving the Euler-Lagrange equations with
appropriate boundary conditions, preserving manifestly the full nonlinearity that is 
characteristic of skyrmion states. It is shown that
the fluctuations stabilize the skyrmion lattice in a region of the phase diagram
close to the critical temperature, where it becomes the equilibrium state.
A comparison of this approach with previous computations performed with a different 
approach (truncated Fourier expansion of magnetic states) is given. 
\end{abstract}

\pacs{111222-k}

\maketitle

%\section{}
%\subsection{}
%\subsubsection{}

\section{Introduction}

Cubic helimagnets support chiral magnetic textures called skyrmions that are very interesting
from the fundamental and applied physics points of view. From the point of view of
fundamental physics, skyrmions are solitonic states that emerge as topologically
nontrivial solutions of nonlinear field equations \cite{Roessler06}.
From the point of view of applied physics, they are very promising ingredients for spintronic 
devices \cite{Fert13,Romming13}, since they are textures modulated 
at the nanoscale, very robust due to the protection provided by the topology,
and appear spontaneously under certain conditions that can be externally
controlled via the temperature, the magnetic field, or the electric current.

Theoretically, skyrmions can appear in bulk materials as isolated excitations of the forced 
ferromagnetic phase (FFM) \cite{Bogdanov94b}. However, mean field computations indicate
that they cannot condense spontaneously into equilibrium states such as a skyrmion lattice
(SKL), since the competing conical state (CS) has lower free energy \cite{Roessler06,Wright89}.
To stabilize the SKL other ingredients, like high magnetic anisotropy \cite{Butenko10} or
modifications of the magnetic stiffness \cite{Roessler06} are necessary.
In two dimensional space, however, the SKL become the equilibrium state in some regions of the
phase diagram \cite{Rybakov13}. Experimentally, SKLs have been found in thin films samples of
some materials \cite{Yu10}.

Despite the theoretical considerations, the A-phase that appears in bulk cubic helimagnets has 
been identified with a SKL \cite{Muehlbauer09,Munzer10,Seki12}.
It has been proposed by M\"uhlbauer \textit{et al}. that it is 
stabilized by fluctuations that modify the mean field computations \cite{Muehlbauer09}. 
This idea is supported by Monte Carlo simulations \cite{Buhrandt13}.
M\"uhlbauer \textit{et al}. addressed the problem by minimizing the appropriate free
energy using a truncated basis of Fourier modes for the magnetic configurations.
Adding the contribution of fluctuations, they found as equilibrium solution a magnetic state
different from the CS and compatible with a SKL. However, it is unclear how this
equilibrium state is related to the magnetic skyrmions of Bogdanov and collaborators
\cite{Bogdanov89,Bogdanov94a,Bogdanov94b}, which
have a highly nonlinear structure, formed by a central core that has a magnetic moment
pointing opposite to the applied field, surronded by a FFM background in which the magnetic
moments point in the diretion of the field.
The Monte Carlo simulations, although valuable since they provide some insight, are far from
being conclusive, as they are plaged of technical problems: first, the physics of the
problem involve two very different scales, the lattice parameters of the underlying crystal and
of the SKL, that cannot be accomodated in a Monte Carlo simulation;
second, topology divides the configuration space into separated sectors which cause the lost
of ergodicity of the Monte Carlo algorithms.

In this paper we explore the idea that fluctuations stabilize the SKL using a different
approach, working in configuration space with the fully nonlinear skyrmions of Bogdanov.
In this way we clarify the nature 
of the SKL and get insight into the physics of the problem, showing that the
equilibrium skyrmion lattice is a crystal of solitons.
This means that the strong chiral fluctuations that characterize the precursor phase of the A-phase
\cite{Pappas09,Bauer13,Pappas17} should be described by the nonlinear skyrmions as elementary 
excitations, rather than by a bunch of weakly coupled Fourier modes.

To close the introduction, it is worth mentioning that very recently a SKL has been found in 
a cubic helimagnet at low temperature for magnetic field in the range close to the transition
to the FFM \cite{Chacon18}. The existence of a stable SKL in this part of the phase diagram,
stabilized by fluctuations, was predicted in Ref. \onlinecite{Laliena17b}.

\section{Theoretical framework \label{sec:theor}}

Consider a cubic helimagnet whose equilibrium properties are described by the partition function
\begin{equation}
\mathcal{Z}=\int [d\vec{m}]\exp(-\beta\mathcal{W})
\end{equation} 
where $\vec{m}$ is the local order parameter, proportional to the local magnetic moment,
and $\mathcal{W}=q_0\int d^3 r\, W$, a Landau-Ginzburg (LG) functional, with
\begin{equation}
W=\frac{1}{2}\partial_i\vec{m}\cdot\partial_i\vec{m}+
q_0\vec{m}\cdot\nabla\times\vec{m}+q_0^2\left(\frac{a}{2}m^2+\frac{1}{4}m^4-\vec{h}\cdot\vec{m}\right).
\label{eq:W}
\end{equation}
The parameter $q_0$ has the dimensions of inverse length and sets the scale for the spatial 
modulation 
of the magnetic configurations. The index $i$ runs over $\{x,y,z\}$, summation over repeated indices 
is understood and $\partial_i$ stands for $\partial/\partial x_i$.
The first and second terms in (\ref{eq:W}) correspond to the ferromagnetic and Dzyaloshinkii-Moriya
exchange interactions, respectively,
and the last term within the brakets is the Zeeeman energy, with $\vec{h}$ proportional to the 
applied magnetic field, which we take along the $z$-axis: $\vec{h}=h\hat{z}$. 
The quadratic and quartic terms in $\vec{m}$ are standard in LG models: $a$ controls the 
temperature variation, and we fix the coefficient of the quartic term to $1/4$. 
This can always be done
by a suitable rescaling of $\vec{m}$ and a redefinition of $\vec{h}$ and $\beta$.
Alternatively, by a rescaling of $\vec{m}$ we could set $\beta=1$. In this case a different
coefficient of the quartic term appears and we would have a more standard form of the LG 
theory. To study the fluctuations, however, we prefer the equivalent form presented here.

It is convenient to work with the dimensionless free energy density 
\begin{equation}
f=-\frac{1}{\beta q_0^3V}\ln\mathcal{Z},
\end{equation}
where $V$ is the volume. 
To evaluate the partition function we assume that it is dominated by the local minima of 
$\mathcal{W}$, which is a good approximation if $\beta$ is large. The local minima are solutions 
of the Euler-Lagrange~(EL) equations, $\delta\mathcal{W}/\delta\vec{m}=0$, that explicitely read
\begin{equation}
\nabla^2\vec{m}-2q_0\nabla\times\vec{m}-q_0^2(a+m^2)\vec{m}+q_0^2\vec{h}=0.
\label{eq:EL}
\end{equation}
Let us denote a generic local minimum by $\vec{m}_0$ and expand $\mathcal{W}$ in powers of 
$\vec{\xi}=\vec{m}-\vec{m}_0$ up to quadratic order:
\begin{equation}
\mathcal{W}=\mathcal{W}(\vec{m}_0)+q_0\int d^3r\,\xi_\alpha K_{\alpha\beta} \xi_\beta + O(\xi^3),
\label{eq:exp}
\end{equation}
where the indices $\alpha$ and $\beta$ run over $\{x,y,z\}$ and
\begin{equation}
K_{\alpha\beta}=\left[-\nabla^2+q_0^2(a+m_0^2)\right]\delta_{\alpha\beta}
+2q_0\epsilon_{\alpha\beta\gamma}\partial_\gamma+2q_0^2 m_{0\alpha} m_{0\beta}.
\end{equation}
is a symmetric differential operator that is define positive since $\vec{m}_0$ is a local minimum 
of $\mathcal{W}$. The symbol $\epsilon_{\alpha\beta\gamma}$ stands for the totally antisymmetric tensor. 
The linear term in~(\ref{eq:exp}) vanishes on account of the EL equations. 
The partition function gets a contribution from each local minimum, obtained as an 
integral over $\xi$ that can be readily evaluated since the integrand is gaussian. 
The free energy density associated to $\vec{m}_0$ has the form $f = f_0 + f_1$, where 
\begin{equation}
f_0 = \frac{1}{q_0^2V}\int d^3r W(\vec{m}_0),
\label{eq:w0} 
\end{equation}
is sometimes called the mean field or classical contribution, and
\begin{equation}
f_1 = \frac{1}{2\beta q_0^3V}\ln\det(KK_0^{-1})
\label{eq:f1}
\end{equation}
is the contribution of the fluctuations to gaussian order.
The operator $K_{0\alpha\beta}=-\nabla^2\delta_{\alpha\beta}$, which does not depend on $\vec{m}_0$,
is introduced merely as a convenient form to normalize the term coming from fluctuations. 
In the continuum limit $f_1$ is divergent and a short distance cut-off, $\Lambda$, has to be
introduced. The crystal lattice, whose latttice parameter is denoted by $a_\mathrm{L}$, provides 
a natural cut-off, $\Lambda=\pi/a_\mathrm{L}$.

To compute the total free energy one has to solve the spectral problem 
$K_{\alpha\beta}\xi_\beta=\lambda\xi_\alpha$. However, away from criticality, $f_1$ is dominated by 
the short distance fluctuations. That is, $\ln\det(KK_0^{-1})$ is dominated by the largest
eigenvalues of $K$, provided it does not have infrarred divergences, as it is the case if
we are outside the critical region. The largest eigenvalues of $K$ correspond to the short
distance modes, and for these $K\sim K_0$, so that we have 
\begin{eqnarray}
\ln\det(KK_0^{-1}) &=& \mathrm{Tr}\ln[1+(K-K_0)K_0^{-1}] \nonumber \\
&\approx& \mathrm{Tr}[(K-K_0)K_0^{-1}].
\end{eqnarray}
The last approximation in the above expression relays on the dominance of short distance
modes, for which $K-K_0\ll K_0$\footnote{This relation has to be understand in terms of matrix 
elements between short distance modes.}. 
It is straightforward to evaluate $\mathrm{Tr}[(K-K_0)K_0^{-1}]$   
using the basis of eigenstates of $K_0$ given by plane waves 
$\vec{\xi}_\alpha=e^{\mathrm{i}\vec{k}\vec{r}}\hat{u}_\alpha$, with the three
polarization vectors satisfying $\hat{u}_\alpha\cdot\hat{u}_\beta=\delta_{\alpha\beta}$. We obtain
\begin{equation}
f_1=c_0\frac{1}{V}\int d^3r\,m_0^2(r),
\label{eq:f1exp}
\end{equation}
where
\begin{equation}
c_0 = \frac{1}{\beta}\frac{5}{4\pi^2}\frac{\Lambda}{q_0}.
\end{equation}
Expression~(\ref{eq:f1exp}) is general for any local minimum, $\vec{m}_0$, of the LG functional.

Notice that, in spite of its appearance, $f_1$ is not merely a renormalization of the
LG $m^2$ term that can be absorbed in a redefinition of $a$.
The local minimum $\vec{m}_0$ has to be computed with the LG functional, with the parameter
$a$ multiplying the $m^2$ term. Then, the gaussian fluctuation around this local minimum
give the $f_1$ contribution to its free energy, which depends on $a$ through $m_0$.

The saddle point expansion is reliable if the coeffcient $c_0$ is small, that is, if $\beta$ is
large. For the cubic helimagnet MnSi $q_0\approx 0.03$ \AA$^{-1}$ and $a_\mathrm{L}\approx 4.5$ \AA,
so that $\Lambda/q_0\approx 20$. Hence we may take $c_0=2.5/\beta$. The value of $\beta$ is 
unknown, but it has to be large to justify the saddle point expansion.
\begin{figure}[t!]
\begin{center}
\includegraphics[width=0.8\linewidth]{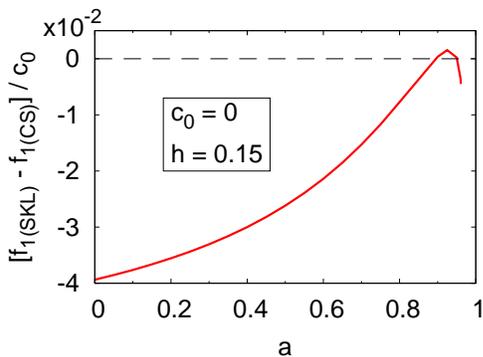}
\caption{Difference in the free energy of fluctuations between the skyrmion lattice and the
conical state computed in the mean field case ($c_0=0$). 
\label{fig:delta_m2}}
\end{center}
\end{figure}

\begin{figure}[t!]
\begin{center}
\includegraphics[width=0.8\linewidth]{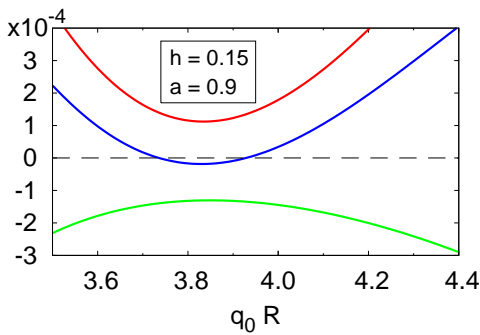}
\caption{Components of the free
energy including gaussian fluctuations as a function of the SKL cell radius, $R$,
for the parameters displayed and with $c_0=0.01$; the blue line is the difference between the
free energies of the SKL and the CS; the red and green lines are the contributions
of the mean field and fluctuations, respectively, to this difference.
\label{fig:fe}}
\end{center}
\end{figure}

The validity of the saddle point expansion and of the short distance approximation is limited by
the behavior of the soft modes. Usually these become important only in a neighborhood of the
critical region, where they produce infrared divergences that invalidate the saddle point
expansion and, of course, the short distance approximation. The fact that the transitions are
of first order \cite{Bauer13,Pappas17} means that it is possible that the soft modes play little 
role in the chiral states, so that that most of the phase diagram obtained in the gaussian 
approximation is qualitatively valid. In any case, within the present approach it is
not possible to delimite the region of validity of the computations, given that we do not
consider the soft modes. 
In addition, our treament of the CS, presented in Sec.~\ref{sec:conical}, is not
valid for very low values of the magnetic field. In this case the CS is nearly
degenerate, as its wave vector can be rotated away from the magnetic field direction with a 
small energy cost. This situation requires a modification of mean field theory known
as Brazovskii theory \cite{Janoschek13,Brazovskii75}.

\section{Conical state \label{sec:conical}}

The CS is a solution of the EL equations of the form 
$\vec{m}_0=(m_\mathrm{t}\cos qz, m_\mathrm{t}\sin qz,m_z)$, where $m_\mathrm{t}$, $m_z$, and $q$ 
are constants. The EL equations are satisfied if and only if
\begin{eqnarray}
m_z &=& \frac{h}{1-\Delta^2}, \\
m_\mathrm{t}^2 &=& 1 - a - \Delta^2 -\frac{h^2}{(1-\Delta^2)^2}, \label{eq:mt}
\end{eqnarray}
where $\Delta = (q-q_0)/q_0$. Notice that $\Delta^2$ is limited by the inequality
$m_\mathrm{t}^2\ge 0$, what implies that $\Delta^2\le\Delta^2_{\mathrm{max}}$, where 
$\Delta^2_{\mathrm{max}}$ satisfy the equation
\begin{equation}
1-a-\Delta^2_{\mathrm{max}} - \frac{h^2}{(1-\Delta^2_{\mathrm{max}})^2} = 0.
\end{equation}
It is not difficult to see that $\Delta^2_{\mathrm{max}}\le 1-a$ and that $\Delta_{\mathrm{max}}=0$
for $a=1-h^2$. For $a>1-h^2$ Eq.~(\ref{eq:mt}) gives $m_\mathrm{t}<0$ for any $\Delta^2$. 
Thus, the CS exists only for $a<1-h^2$.

The free energy density including gaussian fluctuations is
\begin{equation}
f_{\mathrm{C}} = -\frac{(1-a-\Delta^2)^2}{4}-\frac{h^2}{2(1-\Delta^2)}+c_0(1-a-\Delta^2).
\end{equation}
The last term, proportional to $c_0$, is the contribution of the fluctuations in the
short distance approximation, given by Eq.~(\ref{eq:f1exp}). The equilibrium value of
$\Delta$ is obtained by minimizing the free energy. In absence of fluctuations,
$c_0=0$, the minimum of the free energy is attained at $\Delta=0$, that is, $q=q_0$,
for any value of $a$ and $h$. A second order phase transition to the FFM state takes
place when $a=1-h^2$. This is the mean field critical line.
For small $c_0$ the free energy minimum is still attained at $\Delta=0$. 
This minimum disappears on the critical line $a=1-h^2-2c_0$. We see that, as expected,
the fluctuations lower the critical temperature.

\section{Skyrmion lattice}

The EL equations~(\ref{eq:EL}) have axisymmetric solutions of solitonic nature
called skyrmions \cite{Roessler06}.
Using the parametrization of the local order parameter in polar coordinates
\begin{equation}
\vec{m}=m\left(\sin\theta\cos\psi,\sin\theta\sin\psi,\cos\theta\right),
\end{equation}
and cylindric coordinates $(r,\varphi,z)$ for the spatial points, the skyrmion is a solution
in which $\psi=\varphi+\pi/2$ and $m$ and $\theta$ are functions of the radial coordinate, $r$, 
only, so that it is invariant under rotations around the magnetic field axis, $\hat{z}$.
The EL can be cast to the form
\begin{widetext}
\begin{eqnarray}
m^{\prime\prime}+\frac{m^\prime}{r}-m\left[{\theta^\prime}^2+\frac{\sin^2\theta}{r^2}
+2q_0\left(\theta^\prime+\frac{\sin\theta\cos\theta}{r}\right)+a+m^2\right]+h\cos\theta &=0, 
\label{eq:ELm} \\
\theta^{\prime\prime}+\frac{\theta^\prime}{r}+2\frac{m^\prime}{m}(\theta^\prime+q_0)
-\frac{\sin\theta\cos\theta}{r^2}+2q_0\frac{\sin^2\theta}{r}-\frac{h}{m}\sin\theta &=0, 
\label{eq:ELt}
\end{eqnarray}
\end{widetext}
where the prime stands for the derivative with respect to $r$. The boundary conditions are
\begin{equation}
%\begin{array}{ll}
\theta(0) = \pi,  \lim_{r\rightarrow\infty}\theta(r) =0, %\\
m^\prime(0)=0,  \lim_{r\rightarrow\infty} m(r)=m_\mathrm{s},
%\end{array}
\end{equation}
where the non negative constant $m_\mathrm{s}$ is the value of the order parameter in the 
FFM homogeneous state, since as $r\rightarrow\infty$ Eq.~(\ref{eq:ELm}) approaches the EL 
equation for the homogeneous state.
The condition $m^\prime(0)=0$ is necessary to have a finite solution at $r=0$.
Thus, the skyrmion has no degree of freedom. Since $\theta$ and $m$ tend exponentially
to 0 and $m_\mathrm{s}$ as $r\rightarrow\infty$, the skyrmion is a localized structure that consists
of a central core surronded by a FM background\footnote{The approach of $\theta$ and $m$ to their 
limits as $r\rightarrow\infty$ is not monotonous but oscillating, with the oscillations 
exponentially damped.}. The core size is controlled by the external parameters $a$ and $h$.

It is believed that A-phase that appears in several cubic helimagnet is formed by the
condensation of skyrmions in a lattice with hexagonal cells that have a skyrmion core in its center.
The lattice breaks the skyrmion rotational symmetry. However, if the lattice cell is
larger than the skyrmion core, we can approximate the lattice cell by a central axisymmetric skyrmion
core surronded by a deformed FM background. Then, the energy of the skyrmion lattice can be computed
in the so called circular cell approximation (CCA) \cite{Bogdanov94a}, in which the skyrmion profile
is a solution of Eqs.~(\ref{eq:ELm}) and~(\ref{eq:ELt}) on a circle of radius $R$, proportional to the
cell size, with the boundary conditions:
\begin{equation}
\theta(0) = \pi, \quad \theta(R) =0, \quad m^\prime(0)=0, \quad m(R)=m_\mathrm{s}.
\label{eq:BCSKL}
\end{equation}
The skyrmion lattice (SKL) has thus two free parameters,  $R$ and $m_\mathrm{s}$, that are fixed 
by minimizing the free energy. Notice that in this case $m_\mathrm{s}$ is not forced to be equal to
the value of the order parameter in the homogeneous state.

\begin{figure}[t!]
\begin{center}
\includegraphics[width=\linewidth]{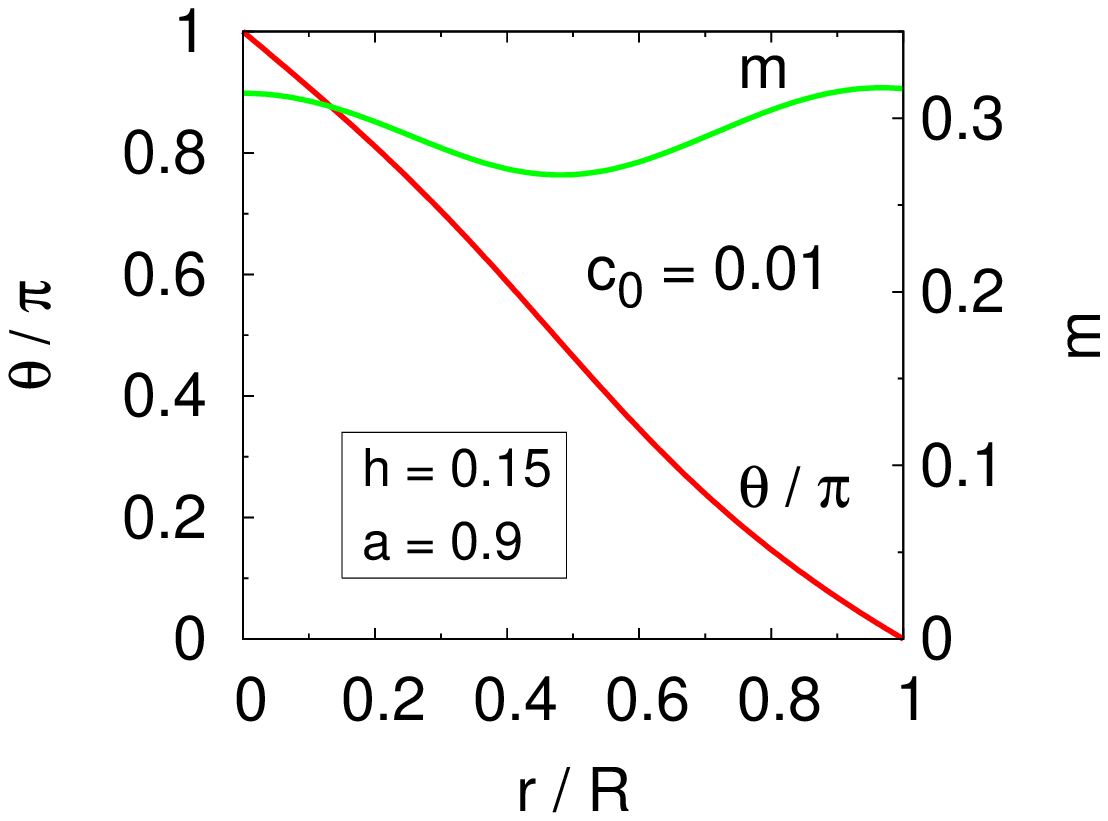}
\caption{Equilibrium skyrmion profile in the unit cell for the parameters displayed in the legend,
including fluctuations.
\label{fig:config}}
\end{center}
\end{figure}

At mean field level, the SKL has higher free energy than the CS for any value of $a$ and $h$;
thus it is at most metastable. To obtain a stable SKL one has to either modify the
model \cite{Roessler06,Butenko10} or go beyond the mean field approximation and include the effect of 
fluctuations at gaussian level \cite{Muehlbauer09}. Let us analyze the last possibility.
Fig.~\ref{fig:delta_m2} displays the difference of $f_1/c_0$, given by Eq.~(\ref{eq:f1exp}),
between the SKL and the CS, computed with the mean field equilibrium solutions.
As this difference is negative except in a very narrow interval of $a$ in the vicinity of the 
transition point, we see that the free energy of fluctuations contributes to lower the free energy 
of the SKL with respect to the CS. Thus, thermal fluctuations are a potential mechanism to 
stabilize the SKL.

\begin{figure}[t!]
\begin{center}
\includegraphics[width=\linewidth]{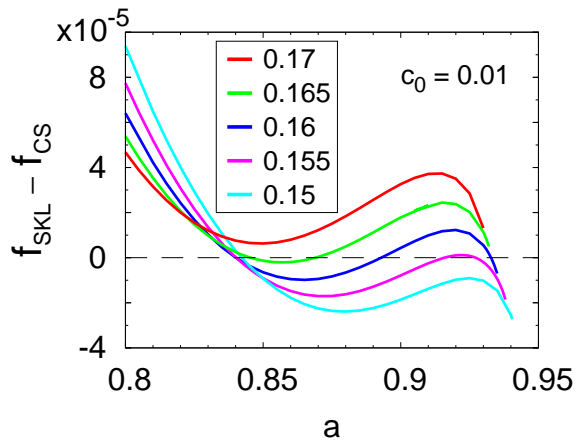}
\caption{Free energy difference between the SKL and CS, $f_{\mathrm{SKL}}-f_\mathrm{C}$,
as a function of $a$ for the fixed values of $h$ displayed in the legend.
\label{fig:fevsa}}
\end{center}
\end{figure}

To explore this possibility, we solve numerically the boundary value problem defined by 
Eqs.~(\ref{eq:ELm}) and~(\ref{eq:ELt}) and the boundary conditions~(\ref{eq:BCSKL}). 
Then, we determine the equilibrium values of $m_\mathrm{s}$ and $R$ by minimizing the free energy 
that includes the fluctuations, 
$f=f_0+f_1$. We fix $c_0=0.01$. With such small value of $c_0$, the equilibrium values of 
$m_\mathrm{s}$ and $R$ are only slightly shifted from their mean field values.
A typical result is shown in Fig.~\ref{fig:fe}, which displays the
components of the free energy as a function of $q_0R$. In these plots the value of $m_\mathrm{s}$
has been fixed by minimizing the total free energy, keeping $R$ fixed. 
The blue line is the difference of total free energies between the SKL and the CS.
The red and green lines represent respectively the mean field ($f_0$) and fluctuation ($f_1$) 
components of the this free energy difference.
The SKL is more stable than the CS since its total free energy is lower in a neighbourhood of the 
minimum. Thus, in this case ($h=0.15$ and $a=0.9$), the SKL is the equilibrium state.
The equilibrim skyrmion profile in the unit cell is displayed in Fig.~(\ref{fig:config}).

\section{Phase diagram and discussion}

Fig.~\ref{fig:fevsa} displays the difference between the free energies of the SKL and CS,
$f_{\mathrm{SKL}}-f_\mathrm{CS}$, as a function of $a$ for fixed values of $h$, with $c_0=0.01$.
The lines end at the point at which the SKL disappears as a metastable state.
Four cases can be distinguised, depending on the range of $h$.
a) For $h\gtrsim 0.167$ the SKL is never the equilibrium state. 
b) For $0.162\lesssim h \lesssim 0.167$, it becomes the equilibrium state in an interval of $a$;
below and above this interval it is metastable or unstable.
c) For $0.155\lesssim h \lesssim 0.162$ the SKL is the equilibrium state in two disjoint intervals 
of $a$: by increasing $a$ from negative values, where the SKL is metastable, it becomes the 
equilibrium state at a certain $a$; at a higher $a$ 
the SKL loses stability and becomes again metastable; and it regains stability at a still higher 
value of $a$, and remains the equilibrium state until it disappears.
d) For $h\lesssim 0.155$, the SKL becomes the equilibrium state at high enough $a$ and remains so
until its disappearance. The phase diagram is reconstructed from this results. Let us discuss it in 
the following paragraphs.

\begin{figure}[t!]
\begin{center}
\includegraphics[width=\linewidth]{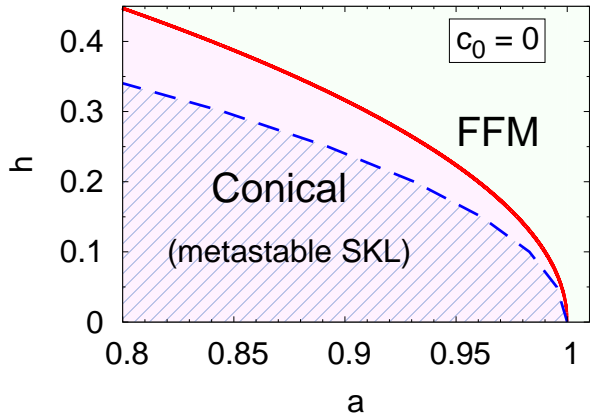}%
\caption{Phase diagram at the mean field level ($c_0=0$). The conical state
is the equilibrium state in the red region. In the blue stripped region the skyrmion lattice
is metastable.
\label{fig:phd_mf}}%
\end{center}
\end{figure}

At mean field level ($c_0=0$), the phase diagram
is displayed in Fig.~\ref{fig:phd_mf}. A second order phase transition
takes place on the red line, which separates the homogeneous FFM phase (green region) from the 
conical phase (red region).
The SKL is metastable within the conical phase, in the region with blue stripes. The local 
minimum of the free energy that defines the SKL disappears on the dashed blue line. In contrast 
with what happens in models where the modulus of the order parameter is fixed, the SKL does not 
disappear through a
nucleation process, in which the lattice size $R$ diverges and the homogeneous state is attained
smoothly \cite{Bogdanov94a,Butenko10}. In the present case, with a soft modulus, the size of the 
SKL remains bounded and the metastable state disappears because the minimum depth gets 
gradually shallower until the minimum becomes a inflection point.

\begin{figure}[t!]
\begin{center}
\includegraphics[width=\linewidth]{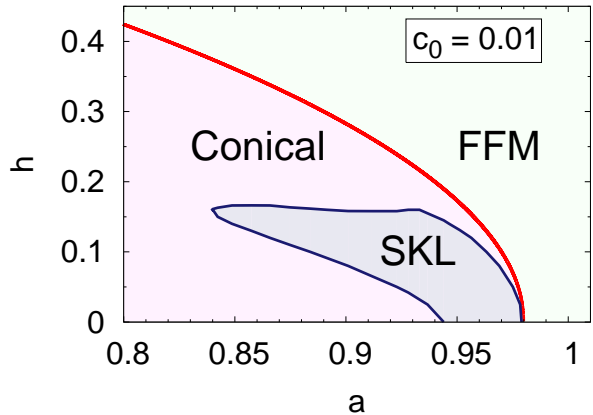}%
\caption{Phase diagram including gaussian fluctuations ($c_0=0.01$). The SKL is the equilibrium
state in the blue region. On the blue line a first order phase transition takes place.
\label{fig:phd}}%
\end{center}
\end{figure}

Figure~\ref{fig:phd} displays the phase diagram including fluctuations, with $c_0=0.01$.
The effect of the fluctuations is quantitative, shifting the line boundary between the homogeneous and
CS to lower values of $a$ (lower temperatures), and qualitative, stabilizing the SKL 
in the blue region. On the phase boundary (blue line), a first order transition
takes place, since the SKL and the CS cannot be smoothly conected.
With higher values of $c_0$ the phase diagram is similar, with the SKL stable phase extended to lower 
values of $a$. For instance, with $c_0=0.1$ the region where the SKL is stable has a similar form
to the blue region of Fig.~\ref{fig:phd} (right), but it covers the interval from $a=-0.4$ to $a=0.75$
and from $h=0$ to $h=0.5$. In this case, the transition from the FFM to the CS
takes place $a=0.8$ for $h=0$.

The results may be invalid near the phase boundary in the high $a$ region, where
soft modes may become important, causing the
failure of both the saddle point expansion and the short distance approximation.
Also, the high (approximate) degeneracy of the CS for low field values invalidates the
computation in the low field region, where the CS has to be treated with a
mean field theory of Brazovskii type.

The phase diagram is very similar to that obtained by M\"uhlbauer \textit{et al.} 
\cite{Muehlbauer09} using a a truncated Fourier basis to obtain the mean field local minima and 
to minimize the total free energy.
These authors identify the new equilibrium state stabilized by the fluctuations with a SKL. 
Although this identification is reasonable, it is unclear how the spin texture determined through
the summation of a finite number of Fourier modes is related to the highly nonlinear skyrmion 
configurations of Bogdanov.
We have shown here that results similar to those of M\"uhlbauer \textit{et al.} can be obtained by
working directly with the fully nonlinear skyrmion texture in configuration space: a lattice formed
by the condensation of skyrmions is the equilibrium state at low enough field near the critical
temperature. It can be identified with the A-phase of cubic helimagnets. These results 
shed light on the internal structure of the SKL and thus provide more insight to the physical aspects
of the problem.
Of special importance is the fact that the strong chiral fluctuations that characterize the precursor 
phase of the A-phase \cite{Bauer13} should be described by nonlinear skyrmion tubes as elementary 
excitations, rather than by the dynamics of a bunch of weakly coupled Fourier modes.

\begin{acknowledgments}
V.L. and J.C. thank A. Bogdanov and A. Leonov for useful discussions.
The authors acknowledge the Grant No. MAT2015-68200- C2-2-P from the Spanish Ministry of 
Economy and Competitiveness. This work was partially supported by the scientific JSPS 
Grant-in-Aid for Scientific Research (S) (Grant No. 25220803), 
and the MEXT program for promoting the enhancement of research universities, 
and JSPS Core-to-Core Program, A. Advanced Research Networks.
\end{acknowledgments}

\bibliographystyle{apsrev4-1}
\bibliography{references}

\begin{thebibliography}{23}%
\makeatletter
\providecommand \@ifxundefined [1]{%
 \@ifx{#1\undefined}
}%
\providecommand \@ifnum [1]{%
 \ifnum #1\expandafter \@firstoftwo
 \else \expandafter \@secondoftwo
 \fi
}%
\providecommand \@ifx [1]{%
 \ifx #1\expandafter \@firstoftwo
 \else \expandafter \@secondoftwo
 \fi
}%
\providecommand \natexlab [1]{#1}%
\providecommand \enquote  [1]{``#1''}%
\providecommand \bibnamefont  [1]{#1}%
\providecommand \bibfnamefont [1]{#1}%
\providecommand \citenamefont [1]{#1}%
\providecommand \href@noop [0]{\@secondoftwo}%
\providecommand \href [0]{\begingroup \@sanitize@url \@href}%
\providecommand \@href[1]{\@@startlink{#1}\@@href}%
\providecommand \@@href[1]{\endgroup#1\@@endlink}%
\providecommand \@sanitize@url [0]{\catcode `\\12\catcode `\$12\catcode
  `\&12\catcode `\#12\catcode `\^12\catcode `\_12\catcode `\%12\relax}%
\providecommand \@@startlink[1]{}%
\providecommand \@@endlink[0]{}%
\providecommand \url  [0]{\begingroup\@sanitize@url \@url }%
\providecommand \@url [1]{\endgroup\@href {#1}{\urlprefix }}%
\providecommand \urlprefix  [0]{URL }%
\providecommand \Eprint [0]{\href }%
\providecommand \doibase [0]{http://dx.doi.org/}%
\providecommand \selectlanguage [0]{\@gobble}%
\providecommand \bibinfo  [0]{\@secondoftwo}%
\providecommand \bibfield  [0]{\@secondoftwo}%
\providecommand \translation [1]{[#1]}%
\providecommand \BibitemOpen [0]{}%
\providecommand \bibitemStop [0]{}%
\providecommand \bibitemNoStop [0]{.\EOS\space}%
\providecommand \EOS [0]{\spacefactor3000\relax}%
\providecommand \BibitemShut  [1]{\csname bibitem#1\endcsname}%
\let\auto@bib@innerbib\@empty
%</preamble>
\bibitem [{\citenamefont {R{\"o\ss}ler}\ \emph {et~al.}(2006)\citenamefont
  {R{\"o\ss}ler}, \citenamefont {Bogdanov},\ and\ \citenamefont
  {Pfleiderer}}]{Roessler06}%
  \BibitemOpen
  \bibfield  {author} {\bibinfo {author} {\bibfnamefont {U.~K.}\ \bibnamefont
  {R{\"o\ss}ler}}, \bibinfo {author} {\bibfnamefont {A.~N.}\ \bibnamefont
  {Bogdanov}}, \ and\ \bibinfo {author} {\bibfnamefont {C.}~\bibnamefont
  {Pfleiderer}},\ }\href@noop {} {\bibfield  {journal} {\bibinfo  {journal}
  {Nature}\ }\textbf {\bibinfo {volume} {442}},\ \bibinfo {pages} {797}
  (\bibinfo {year} {2006})}\BibitemShut {NoStop}%
\bibitem [{\citenamefont {Fert}\ \emph {et~al.}(2013)\citenamefont {Fert},
  \citenamefont {Cross},\ and\ \citenamefont {Sampaio}}]{Fert13}%
  \BibitemOpen
  \bibfield  {author} {\bibinfo {author} {\bibfnamefont {A.}~\bibnamefont
  {Fert}}, \bibinfo {author} {\bibfnamefont {V.}~\bibnamefont {Cross}}, \ and\
  \bibinfo {author} {\bibfnamefont {J.}~\bibnamefont {Sampaio}},\ }\href@noop
  {} {\bibfield  {journal} {\bibinfo  {journal} {Nature Nanotechnology}\
  }\textbf {\bibinfo {volume} {8}},\ \bibinfo {pages} {152} (\bibinfo {year}
  {2013})}\BibitemShut {NoStop}%
\bibitem [{\citenamefont {Romming}\ \emph {et~al.}(2001)\citenamefont
  {Romming}, \citenamefont {Hanneken}, \citenamefont {Menzel}, \citenamefont
  {Bickel}, \citenamefont {Wolter}, \citenamefont {von Bergmann}, \citenamefont
  {Kubetzka},\ and\ \citenamefont {Wiesendanger}}]{Romming13}%
  \BibitemOpen
  \bibfield  {author} {\bibinfo {author} {\bibfnamefont {N.}~\bibnamefont
  {Romming}}, \bibinfo {author} {\bibfnamefont {C.}~\bibnamefont {Hanneken}},
  \bibinfo {author} {\bibfnamefont {M.}~\bibnamefont {Menzel}}, \bibinfo
  {author} {\bibfnamefont {J.}~\bibnamefont {Bickel}}, \bibinfo {author}
  {\bibfnamefont {B.}~\bibnamefont {Wolter}}, \bibinfo {author} {\bibfnamefont
  {K.}~\bibnamefont {von Bergmann}}, \bibinfo {author} {\bibfnamefont
  {A.}~\bibnamefont {Kubetzka}}, \ and\ \bibinfo {author} {\bibfnamefont
  {R.}~\bibnamefont {Wiesendanger}},\ }\href@noop {} {\bibfield  {journal}
  {\bibinfo  {journal} {Science}\ }\textbf {\bibinfo {volume} {341}},\ \bibinfo
  {pages} {636} (\bibinfo {year} {2001})}\BibitemShut {NoStop}%
\bibitem [{\citenamefont {Bogdanov}\ and\ \citenamefont
  {Hubert}(1994{\natexlab{a}})}]{Bogdanov94b}%
  \BibitemOpen
  \bibfield  {author} {\bibinfo {author} {\bibfnamefont {A.}~\bibnamefont
  {Bogdanov}}\ and\ \bibinfo {author} {\bibfnamefont {A.}~\bibnamefont
  {Hubert}},\ }\href@noop {} {\bibfield  {journal} {\bibinfo  {journal} {phys.
  stat. sol. (b)}\ }\textbf {\bibinfo {volume} {186}},\ \bibinfo {pages} {527}
  (\bibinfo {year} {1994}{\natexlab{a}})}\BibitemShut {NoStop}%
\bibitem [{\citenamefont {Wright}\ and\ \citenamefont
  {Mermin}(1989)}]{Wright89}%
  \BibitemOpen
  \bibfield  {author} {\bibinfo {author} {\bibfnamefont {D.~C.}\ \bibnamefont
  {Wright}}\ and\ \bibinfo {author} {\bibfnamefont {N.~D.}\ \bibnamefont
  {Mermin}},\ }\href@noop {} {\bibfield  {journal} {\bibinfo  {journal}
  {Reviews of Modern Physics}\ }\textbf {\bibinfo {volume} {61}},\ \bibinfo
  {pages} {385} (\bibinfo {year} {1989})}\BibitemShut {NoStop}%
\bibitem [{\citenamefont {Butenko}\ \emph {et~al.}(2010)\citenamefont
  {Butenko}, \citenamefont {Leonov}, \citenamefont {R{\"o\ss}ler},\ and\
  \citenamefont {Bogdanov}}]{Butenko10}%
  \BibitemOpen
  \bibfield  {author} {\bibinfo {author} {\bibfnamefont {A.~B.}\ \bibnamefont
  {Butenko}}, \bibinfo {author} {\bibfnamefont {A.~A.}\ \bibnamefont {Leonov}},
  \bibinfo {author} {\bibfnamefont {U.~K.}\ \bibnamefont {R{\"o\ss}ler}}, \
  and\ \bibinfo {author} {\bibfnamefont {A.~N.}\ \bibnamefont {Bogdanov}},\
  }\href@noop {} {\bibfield  {journal} {\bibinfo  {journal} {Physical Review
  B}\ }\textbf {\bibinfo {volume} {82}},\ \bibinfo {pages} {052403} (\bibinfo
  {year} {2010})}\BibitemShut {NoStop}%
\bibitem [{\citenamefont {Rybakov}\ \emph {et~al.}(2013)\citenamefont
  {Rybakov}, \citenamefont {Borisov},\ and\ \citenamefont
  {Bogdanov}}]{Rybakov13}%
  \BibitemOpen
  \bibfield  {author} {\bibinfo {author} {\bibfnamefont {F.}~\bibnamefont
  {Rybakov}}, \bibinfo {author} {\bibfnamefont {A.}~\bibnamefont {Borisov}}, \
  and\ \bibinfo {author} {\bibfnamefont {A.}~\bibnamefont {Bogdanov}},\
  }\href@noop {} {\bibfield  {journal} {\bibinfo  {journal} {Physical Review
  B}\ }\textbf {\bibinfo {volume} {87}},\ \bibinfo {pages} {094424} (\bibinfo
  {year} {2013})}\BibitemShut {NoStop}%
\bibitem [{\citenamefont {Yu}\ \emph {et~al.}(2010)\citenamefont {Yu},
  \citenamefont {Onose}, \citenamefont {Kanazawa}, \citenamefont {Park},
  \citenamefont {Han}, \citenamefont {Matsui}, \citenamefont {Nagaosa},\ and\
  \citenamefont {Tokura}}]{Yu10}%
  \BibitemOpen
  \bibfield  {author} {\bibinfo {author} {\bibfnamefont {X.}~\bibnamefont
  {Yu}}, \bibinfo {author} {\bibfnamefont {Y.}~\bibnamefont {Onose}}, \bibinfo
  {author} {\bibfnamefont {N.}~\bibnamefont {Kanazawa}}, \bibinfo {author}
  {\bibfnamefont {J.}~\bibnamefont {Park}}, \bibinfo {author} {\bibfnamefont
  {J.}~\bibnamefont {Han}}, \bibinfo {author} {\bibfnamefont {Y.}~\bibnamefont
  {Matsui}}, \bibinfo {author} {\bibfnamefont {N.}~\bibnamefont {Nagaosa}}, \
  and\ \bibinfo {author} {\bibfnamefont {Y.}~\bibnamefont {Tokura}},\
  }\href@noop {} {\bibfield  {journal} {\bibinfo  {journal} {Nature}\ }\textbf
  {\bibinfo {volume} {465}},\ \bibinfo {pages} {901} (\bibinfo {year}
  {2010})}\BibitemShut {NoStop}%
\bibitem [{\citenamefont {M{\"u}hlbauer}\ \emph {et~al.}(2009)\citenamefont
  {M{\"u}hlbauer}, \citenamefont {Binz}, \citenamefont {Jonietz}, \citenamefont
  {Pfleiderer}, \citenamefont {Rosch}, \citenamefont {Neubauer}, \citenamefont
  {Georgii},\ and\ \citenamefont {B{\"o}ni}}]{Muehlbauer09}%
  \BibitemOpen
  \bibfield  {author} {\bibinfo {author} {\bibfnamefont {S.}~\bibnamefont
  {M{\"u}hlbauer}}, \bibinfo {author} {\bibfnamefont {B.}~\bibnamefont {Binz}},
  \bibinfo {author} {\bibfnamefont {F.}~\bibnamefont {Jonietz}}, \bibinfo
  {author} {\bibfnamefont {C.}~\bibnamefont {Pfleiderer}}, \bibinfo {author}
  {\bibfnamefont {A.}~\bibnamefont {Rosch}}, \bibinfo {author} {\bibfnamefont
  {A.}~\bibnamefont {Neubauer}}, \bibinfo {author} {\bibfnamefont
  {R.}~\bibnamefont {Georgii}}, \ and\ \bibinfo {author} {\bibfnamefont
  {P.}~\bibnamefont {B{\"o}ni}},\ }\href@noop {} {\bibfield  {journal}
  {\bibinfo  {journal} {Science}\ }\textbf {\bibinfo {volume} {323}},\ \bibinfo
  {pages} {915} (\bibinfo {year} {2009})}\BibitemShut {NoStop}%
\bibitem [{\citenamefont {M{\"u}nzer}\ \emph {et~al.}(2010)\citenamefont
  {M{\"u}nzer}, \citenamefont {Neubauer}, \citenamefont {Adams}, \citenamefont
  {M{\"u}hlbauer}, \citenamefont {Franz}, \citenamefont {Jonietz},
  \citenamefont {Georgii}, \citenamefont {B{\"o}ni}, \citenamefont {Pedersen},
  \citenamefont {Schmidt}, \citenamefont {Rosch},\ and\ \citenamefont
  {Pfleiderer}}]{Munzer10}%
  \BibitemOpen
  \bibfield  {author} {\bibinfo {author} {\bibfnamefont {W.}~\bibnamefont
  {M{\"u}nzer}}, \bibinfo {author} {\bibfnamefont {A.}~\bibnamefont
  {Neubauer}}, \bibinfo {author} {\bibfnamefont {T.}~\bibnamefont {Adams}},
  \bibinfo {author} {\bibfnamefont {S.}~\bibnamefont {M{\"u}hlbauer}}, \bibinfo
  {author} {\bibfnamefont {C.}~\bibnamefont {Franz}}, \bibinfo {author}
  {\bibfnamefont {F.}~\bibnamefont {Jonietz}}, \bibinfo {author} {\bibfnamefont
  {R.}~\bibnamefont {Georgii}}, \bibinfo {author} {\bibfnamefont
  {P.}~\bibnamefont {B{\"o}ni}}, \bibinfo {author} {\bibfnamefont
  {B.}~\bibnamefont {Pedersen}}, \bibinfo {author} {\bibfnamefont
  {M.}~\bibnamefont {Schmidt}}, \bibinfo {author} {\bibfnamefont
  {A.}~\bibnamefont {Rosch}}, \ and\ \bibinfo {author} {\bibfnamefont
  {C.}~\bibnamefont {Pfleiderer}},\ }\href@noop {} {\bibfield  {journal}
  {\bibinfo  {journal} {Physical Review B}\ }\textbf {\bibinfo {volume} {81}},\
  \bibinfo {pages} {041203(R)} (\bibinfo {year} {2010})}\BibitemShut {NoStop}%
\bibitem [{\citenamefont {Seki}\ \emph {et~al.}(2012)\citenamefont {Seki},
  \citenamefont {Yu},\ and\ \citenamefont {Tokura}}]{Seki12}%
  \BibitemOpen
  \bibfield  {author} {\bibinfo {author} {\bibfnamefont {S.}~\bibnamefont
  {Seki}}, \bibinfo {author} {\bibfnamefont {X.~Z.}\ \bibnamefont {Yu}}, \ and\
  \bibinfo {author} {\bibfnamefont {Y.}~\bibnamefont {Tokura}},\ }\href
  {\doibase 10.1126/science.1214143} {\bibfield  {journal} {\bibinfo  {journal}
  {Science}\ }\textbf {\bibinfo {volume} {336}},\ \bibinfo {pages} {198}
  (\bibinfo {year} {2012})}\BibitemShut {NoStop}%
\bibitem [{\citenamefont {Buhrandt}\ and\ \citenamefont
  {Fritz}(2013)}]{Buhrandt13}%
  \BibitemOpen
  \bibfield  {author} {\bibinfo {author} {\bibfnamefont {S.}~\bibnamefont
  {Buhrandt}}\ and\ \bibinfo {author} {\bibfnamefont {L.}~\bibnamefont
  {Fritz}},\ }\href@noop {} {\bibfield  {journal} {\bibinfo  {journal}
  {Physical Review B}\ }\textbf {\bibinfo {volume} {88}},\ \bibinfo {pages}
  {195137} (\bibinfo {year} {2013})}\BibitemShut {NoStop}%
\bibitem [{\citenamefont {Bogdanov}\ and\ \citenamefont
  {Yablonskii}(1989)}]{Bogdanov89}%
  \BibitemOpen
  \bibfield  {author} {\bibinfo {author} {\bibfnamefont {A.~N.}\ \bibnamefont
  {Bogdanov}}\ and\ \bibinfo {author} {\bibfnamefont {D.~A.}\ \bibnamefont
  {Yablonskii}},\ }\href@noop {} {\bibfield  {journal} {\bibinfo  {journal}
  {Sov. Phys. JETP}\ }\textbf {\bibinfo {volume} {68}},\ \bibinfo {pages} {101}
  (\bibinfo {year} {1989})}\BibitemShut {NoStop}%
\bibitem [{\citenamefont {Bogdanov}\ and\ \citenamefont
  {Hubert}(1994{\natexlab{b}})}]{Bogdanov94a}%
  \BibitemOpen
  \bibfield  {author} {\bibinfo {author} {\bibfnamefont {A.}~\bibnamefont
  {Bogdanov}}\ and\ \bibinfo {author} {\bibfnamefont {A.}~\bibnamefont
  {Hubert}},\ }\href@noop {} {\bibfield  {journal} {\bibinfo  {journal}
  {Journal of Magnetism and Magnetic Materials}\ }\textbf {\bibinfo {volume}
  {138}},\ \bibinfo {pages} {255} (\bibinfo {year}
  {1994}{\natexlab{b}})}\BibitemShut {NoStop}%
\bibitem [{\citenamefont {Pappas}\ \emph {et~al.}(2009)\citenamefont {Pappas},
  \citenamefont {Leli{\`e}vre-Berna}, \citenamefont {Falus}, \citenamefont
  {Bentley}, \citenamefont {Moskvin}, \citenamefont {Grigoriev}, \citenamefont
  {Fouquet},\ and\ \citenamefont {Farago}}]{Pappas09}%
  \BibitemOpen
  \bibfield  {author} {\bibinfo {author} {\bibfnamefont {C.}~\bibnamefont
  {Pappas}}, \bibinfo {author} {\bibfnamefont {E.}~\bibnamefont
  {Leli{\`e}vre-Berna}}, \bibinfo {author} {\bibfnamefont {P.}~\bibnamefont
  {Falus}}, \bibinfo {author} {\bibfnamefont {P.~M.}\ \bibnamefont {Bentley}},
  \bibinfo {author} {\bibfnamefont {E.}~\bibnamefont {Moskvin}}, \bibinfo
  {author} {\bibfnamefont {S.}~\bibnamefont {Grigoriev}}, \bibinfo {author}
  {\bibfnamefont {P.}~\bibnamefont {Fouquet}}, \ and\ \bibinfo {author}
  {\bibfnamefont {B.}~\bibnamefont {Farago}},\ }\href@noop {} {\bibfield
  {journal} {\bibinfo  {journal} {Physical Review Letters}\ }\textbf {\bibinfo
  {volume} {102}},\ \bibinfo {pages} {197202} (\bibinfo {year}
  {2009})}\BibitemShut {NoStop}%
\bibitem [{\citenamefont {Bauer}\ \emph {et~al.}(2013)\citenamefont {Bauer},
  \citenamefont {Garst},\ and\ \citenamefont {Pfleiderer}}]{Bauer13}%
  \BibitemOpen
  \bibfield  {author} {\bibinfo {author} {\bibfnamefont {A.}~\bibnamefont
  {Bauer}}, \bibinfo {author} {\bibfnamefont {M.}~\bibnamefont {Garst}}, \ and\
  \bibinfo {author} {\bibfnamefont {C.}~\bibnamefont {Pfleiderer}},\
  }\href@noop {} {\bibfield  {journal} {\bibinfo  {journal} {Physical Review
  Letters}\ }\textbf {\bibinfo {volume} {110}},\ \bibinfo {pages} {177207}
  (\bibinfo {year} {2013})}\BibitemShut {NoStop}%
\bibitem [{\citenamefont {Pappas}\ \emph {et~al.}(2017)\citenamefont {Pappas},
  \citenamefont {Bannenberg}, \citenamefont {Leli{\`e}vre-Berna}, \citenamefont
  {Qian}, \citenamefont {Dewhurst}, \citenamefont {Dalgliesh}, \citenamefont
  {Schlagel}, \citenamefont {Lograsso},\ and\ \citenamefont
  {Falus}}]{Pappas17}%
  \BibitemOpen
  \bibfield  {author} {\bibinfo {author} {\bibfnamefont {C.}~\bibnamefont
  {Pappas}}, \bibinfo {author} {\bibfnamefont {L.}~\bibnamefont {Bannenberg}},
  \bibinfo {author} {\bibfnamefont {E.}~\bibnamefont {Leli{\`e}vre-Berna}},
  \bibinfo {author} {\bibfnamefont {F.}~\bibnamefont {Qian}}, \bibinfo {author}
  {\bibfnamefont {C.}~\bibnamefont {Dewhurst}}, \bibinfo {author}
  {\bibfnamefont {R.}~\bibnamefont {Dalgliesh}}, \bibinfo {author}
  {\bibfnamefont {D.}~\bibnamefont {Schlagel}}, \bibinfo {author}
  {\bibfnamefont {T.}~\bibnamefont {Lograsso}}, \ and\ \bibinfo {author}
  {\bibfnamefont {P.}~\bibnamefont {Falus}},\ }\href@noop {} {\bibfield
  {journal} {\bibinfo  {journal} {Physical Review Letters}\ }\textbf {\bibinfo
  {volume} {119}},\ \bibinfo {pages} {047203} (\bibinfo {year}
  {2017})}\BibitemShut {NoStop}%
\bibitem [{\citenamefont {Chacon}\ \emph {et~al.}(2018)\citenamefont {Chacon},
  \citenamefont {Heinen}, \citenamefont {Halder}, \citenamefont {Bauer},
  \citenamefont {Simeth}, \citenamefont {M{\"u}hlbauer}, \citenamefont
  {Berger}, \citenamefont {Garst}, \citenamefont {Rosch},\ and\ \citenamefont
  {Pfleiderer}}]{Chacon18}%
  \BibitemOpen
  \bibfield  {author} {\bibinfo {author} {\bibfnamefont {A.}~\bibnamefont
  {Chacon}}, \bibinfo {author} {\bibfnamefont {L.}~\bibnamefont {Heinen}},
  \bibinfo {author} {\bibfnamefont {M.}~\bibnamefont {Halder}}, \bibinfo
  {author} {\bibfnamefont {A.}~\bibnamefont {Bauer}}, \bibinfo {author}
  {\bibfnamefont {W.}~\bibnamefont {Simeth}}, \bibinfo {author} {\bibfnamefont
  {S.}~\bibnamefont {M{\"u}hlbauer}}, \bibinfo {author} {\bibfnamefont
  {H.}~\bibnamefont {Berger}}, \bibinfo {author} {\bibfnamefont
  {M.}~\bibnamefont {Garst}}, \bibinfo {author} {\bibfnamefont
  {A.}~\bibnamefont {Rosch}}, \ and\ \bibinfo {author} {\bibfnamefont
  {C.}~\bibnamefont {Pfleiderer}},\ }\href@noop {} {\bibfield  {journal}
  {\bibinfo  {journal} {Nature Physics}\ }\textbf {\bibinfo {volume} {June}}
  (\bibinfo {year} {2018})}\BibitemShut {NoStop}%
\bibitem [{\citenamefont {Laliena}\ and\ \citenamefont
  {Campo}(2017)}]{Laliena17b}%
  \BibitemOpen
  \bibfield  {author} {\bibinfo {author} {\bibfnamefont {V.}~\bibnamefont
  {Laliena}}\ and\ \bibinfo {author} {\bibfnamefont {J.}~\bibnamefont
  {Campo}},\ }\href@noop {} {\bibfield  {journal} {\bibinfo  {journal}
  {Physical Review B}\ }\textbf {\bibinfo {volume} {96}},\ \bibinfo {pages}
  {134420} (\bibinfo {year} {2017})}\BibitemShut {NoStop}%
\bibitem [{Note1()}]{Note1}%
  \BibitemOpen
  \bibinfo {note} {This relation has to be understand in terms of matrix
  elements between short distance modes.}\BibitemShut {Stop}%
\bibitem [{\citenamefont {Janoschek}\ \emph {et~al.}(2013)\citenamefont
  {Janoschek}, \citenamefont {Garst}, \citenamefont {Bauer}, \citenamefont
  {Krautscheid}, \citenamefont {Georgii}, \citenamefont {B{\"o}ni},\ and\
  \citenamefont {Pfleiderer}}]{Janoschek13}%
  \BibitemOpen
  \bibfield  {author} {\bibinfo {author} {\bibfnamefont {M.}~\bibnamefont
  {Janoschek}}, \bibinfo {author} {\bibfnamefont {M.}~\bibnamefont {Garst}},
  \bibinfo {author} {\bibfnamefont {A.}~\bibnamefont {Bauer}}, \bibinfo
  {author} {\bibfnamefont {P.}~\bibnamefont {Krautscheid}}, \bibinfo {author}
  {\bibfnamefont {R.}~\bibnamefont {Georgii}}, \bibinfo {author} {\bibfnamefont
  {P.}~\bibnamefont {B{\"o}ni}}, \ and\ \bibinfo {author} {\bibfnamefont
  {C.}~\bibnamefont {Pfleiderer}},\ }\href@noop {} {\bibfield  {journal}
  {\bibinfo  {journal} {Physical Review B}\ }\textbf {\bibinfo {volume} {87}},\
  \bibinfo {pages} {134407} (\bibinfo {year} {2013})}\BibitemShut {NoStop}%
\bibitem [{\citenamefont {Brazovskii}(1975)}]{Brazovskii75}%
  \BibitemOpen
  \bibfield  {author} {\bibinfo {author} {\bibfnamefont {S.~A.}\ \bibnamefont
  {Brazovskii}},\ }\href@noop {} {\bibfield  {journal} {\bibinfo  {journal}
  {JETP}\ }\textbf {\bibinfo {volume} {41}},\ \bibinfo {pages} {85} (\bibinfo
  {year} {1975})}\BibitemShut {NoStop}%
\bibitem [{Note2()}]{Note2}%
  \BibitemOpen
  \bibinfo {note} {The approach of $\theta $ and $m$ to their limits as
  $r\rightarrow \infty $ is not monotonous but oscillating, with the
  oscillations exponentially damped.}\BibitemShut {Stop}%
\end{thebibliography}%

\end{document}